\documentstyle[12pt,a4wide]{article}
\begin{document}
\title{\bf Light propagation around a relativistic 
vortex flow of dielectric medium}
\author{B. Linet \thanks{E-mail: linet@celfi.phys.univ-tours.fr} \\ 
\small Laboratoire de Math\'ematiques et Physique Th\'eorique \\
\small CNRS/UPRES-A 6083, Universit\'e Fran\c{c}ois Rabelais \\
\small Parc de Grandmont 37200 TOURS, France}
\date{}
\maketitle
\thispagestyle{empty}

\begin{abstract}

We determine the path of the light around a dielectric vortex
described by the relativistic vortex flow of a perfect fluid.

\end{abstract}

\section{Introduction}

Recently, Leonhardt and Piwnicki \cite{le1} have studied the light propagation 
in a vortex flow of non-dispersive dielectric medium in order to obtain
a so-called optical black hole. However, Visser \cite{vi} has shown, in a 
comment of another paper of Leonhardt and Piwnicki \cite{le2}, that this 
vortex geometry has not the sense of black hole such that it is defined in 
general relativity. Moreover, we emphasize that their vortex flow is simply 
an ad hoc generalization of the Newtonian case. Despite these points,
the analysis of  
the light propagation in a dielectric vortex is interesting in principle.

In this work, we take up again the question of the light propagation 
around a dielectric vortex in the case where the relativistic
vortex flow is really determined within the framework of relativistic
hydrodynamics. It is different from the vortex flow considered by
the authors. Then, we establish the qualitative features of the trajectories 
of the light rays dependent on the values of the angular momentum
of the light rays.

The plan of the work is as follows. In section 2, we derive the relativistic
vortex flow. For this case, we calculate in section 3 the effective 
metric for which the light rays define a null congruence of geodesics.
In section 4, we obtain the equations determining the path of the light. 
Then, we discuss the qualitative features of the trajectories 
of the light rays in section 5. 
We add some concluding remarks in section 6.

\section{Relativistic vortex flow}

We are concerned with a perfect fluid having the energy-momentum tensor
\begin{equation}\label{1}
T^{\mu \nu}=(\rho c^2+p)u^{\mu}u^{\nu}+p\eta^{\mu \nu}
\end{equation}
where $\rho$ is the mass density, $p$ the pressure, $u^{\mu}$ the
four-velocity of the fluid, and $\eta_{\mu \nu}$ is the Minkowskian
metric with signature $(-+++)$. It is conserved $\partial_{\mu}T^{\mu \nu}=0$.
We assume the existence of a conserved mass density $r$, i.e. 
$\partial_{\mu}(ru^{\mu})=0$. We also define the relativistic specific 
enthalpy $h=(\rho c^2+p)/r$.

We suppose that the motion of the fluid is isentropic. So, it is not
necessary to consider the specific entropy and we postulate merely an equation
of state $h(p)$. From the equations of the fluid motion, it is
easy to prove that $dh/dp=1/r$. Moreover, by setting
\begin{equation}\label{2}
\mu^{\tau}=hu^{\tau} ,
\end{equation}
we obtain the following equation
\begin{equation}\label{3}
u^{\tau}\partial_{\tau}\mu_{\sigma}=-\partial_{\sigma}h .
\end{equation}

We describe the Minkowskian spacetime in cylindrical coordinates
$(x^0,\rho ,\varphi ,z)$ in which the Minkowskian metric has the form
\begin{equation}\label{4}
ds^2=-(dx^0)^2+d\rho^2+\rho ^2d\varphi^2+dz^2 .
\end{equation}
The symmetries of the vortex flow are given by the three Killing vectors :
$\partial_0$, $\partial_{\varphi}$, $\partial_z$, denoted 
$\xi_{(a)}^{\tau}$ $(a=1,2,3)$, since we assume that the solutions to the 
equations of the fluid motion are stationary and cylindrically symmetric.
In particular, we have 
$\xi_{(a)}^{\tau}\partial_{\tau}h=0$. Taking into account the fact
that $\partial_{\mu}\xi_{(a)\nu}+\partial_{\nu}\xi_{(a)\mu}=0$, we
obtain from (\ref{3}) that
$u^{\sigma}\partial_{\sigma}(\xi_{(a)}^{\tau}\mu_{\tau})=0$. Thus, 
$\xi_{(a)}^{\tau}\mu_{\tau}$ are constants of the fluid motion.
We limit ourselves to the case $\mu^{\rho}=0$ and we put
\begin{equation}\label{6}
\mu_0=-{\cal E}\; ,\; \mu_{\rho}=0\; , \; \mu_{\varphi}={\cal M} \; 
{\rm and} \; \mu_z=k
\end{equation}
where ${\cal E}$ $({\cal E}>0)$, ${\cal M}$ and $k$ are arbitrary 
constants of the
fluid motion. By using a Lorentz transformation
along $z$, we may take $k=0$.

We now solve the equations of the fluid motion. From (\ref{2}), we get 
$h^2=-\mu_{\tau}\mu^{\tau}$. Hence by assumptions (\ref{6}), we find
\begin{equation}\label{7}
h(\rho )=\sqrt{{\cal E}^2-\frac{{\cal M}^2}{\rho^2}}
\end{equation}
which is well defined for $\rho >\rho_c$ where 
$\rho_c=\mid {\cal M}/{\cal E}\mid$. We notice that $h(\rho_c)=0$.
By combining (\ref{2}) and (\ref{7}), we get the components of the 
four-velocity of the fluid
\begin{equation}\label{8}
u^0=\frac{1}{\sqrt{1-\rho_{c}^{2}/\rho^2}}\;, \; 
u^{\rho}=0 \; , \;
u^{\varphi}=\pm \frac{\rho_c}{\rho^2\sqrt{1-\rho_{c}^{2}/\rho^2}} \;
{\rm and}\; u^z=0 \quad {\rm for}\quad \rho >\rho_c  
\end{equation}
in which we choose the sign $+$ henceforth.
Solution (\ref{8}) to the equations of the fluid motion 
gives a model of vortex flow which is only
considered for $\rho >\rho_c$, i.e. outside the core region of 
the vortex flow.
This vortex geometry is usually used in relativistic hydrodynamics 
\cite{bo} and in the relativistic theory of superfluid \cite{ca}.  

\section{Light propagation in a non-dispersive medium}

The electromagnetic properties of an isotrope  non-dispersive dielectric medium
are characterized by the permittivity $\epsilon$ and the permeability
$\mu$. A relativistic model of such a medium has been found a long time ago
\cite{go,ph}. 
The Maxwell equations are written with the aid of a tensor 
$\tilde{g}_{\mu \nu}$ defined by
\begin{equation}\label{9}
\tilde{g}_{\alpha \beta}=\eta_{\alpha \beta}-\left( -1+\frac{1}{n^2}\right)
u_{\alpha}u_{\beta}
\end{equation}
where $n$ is the refractive index, $n^2=\epsilon \mu c^2$, and $u^{\mu}$ the
four-velocity of the medium. 
At the approximation of geometrical optics in this medium, the vector
of propagation of light is $k_{\mu}=\partial_{\mu}S$ where $S$ is the phase 
and it satisfies the eikonal equation \cite{to}
\begin{equation}\label{10}
\tilde{g}^{\alpha \beta}k_{\alpha}k_{\beta}=0
\end{equation}
where $\tilde{g}^{\alpha \beta}$ is the inverse matrix of 
$\tilde{g}_{\mu \nu}$, having from (\ref{9}) the expression
\begin{equation}\label{11}
\tilde{g}^{\alpha \beta}=\eta^{\alpha \beta}-\left( -1+n^2\right) 
u^{\alpha}u^{\beta} .
\end{equation}

Equation (\ref{10}) for the propagation of light in this dielectric medium
is equivalent to the propagation of light in the effective metric 
$\tilde{g}_{\mu \nu}$. We therefore define the components 
$\tilde{k}^{\mu}$ of the vector of propagation by
\begin{equation}\label{12}
\tilde{k}^{\mu}=\tilde{g}^{\mu \nu}k_{\nu}
\end{equation}
and we take $\tilde{k}_{\mu}=k_{\mu}$.
Since $\tilde{k}_{\mu}=\partial_{\mu}S$, equation (\ref{10}) implies
that the vector of propagation $\tilde{k}^{\mu}$ defines a null 
congruence of geodesics for the effective metric 
$\tilde{g}_{\mu \nu}$. We have thereby
\begin{equation}\label{13}
\tilde{k}^{\mu}\tilde{\nabla}_{\mu}\tilde{k}^{\nu}=0
\end{equation}
where $\tilde{\nabla}$ is the metric connection associated to 
$\tilde{g}_{\mu \nu}$.

We suppose that the dielectric medium consists of a perfect fluid.
In the case of the vortex flow given by (\ref{8}), we can calculate
the components of the effective metric
\begin{eqnarray}\label{14}
& & \tilde{g}_{00}=-\frac{1-n^2\rho_{c}^{2}/\rho^2}
{n^2\left( 1-\rho_{c}^{2}/\rho^2\right)} \; , \; 
\tilde{g}_{0\varphi}=-\frac{\rho_c\left( 1-1/n^2\right)}
{1-\rho_{c}^{2}/\rho^2} \; ,\;  
\tilde{g}_{\varphi \varphi}=
\frac{\rho^2\left( 1-\rho_{c}^{2}/n^2\rho^2\right)}
{1-\rho_{c}^{2}/\rho^2} \; , \\ 
& & \nonumber \tilde{g}_{\rho \rho}=1 \; {\rm and}\;  \tilde{g}_{zz}=1
\end{eqnarray}
and these of the inverse metric
\begin{eqnarray}\label{15}
& & \tilde{g}^{00}=-\frac{n^2-\rho_{c}^{2}/\rho^2}
{1-\rho_{c}^{2}/\rho^2} \; , \;
\tilde{g}^{0\varphi}=-\frac{\rho_c(-1+n^2)}{\rho^2\left( 1-\rho_{c}^{2}/
\rho^2\right)} \; , \; 
\tilde{g}^{\varphi \varphi}=\frac{1-n^2\rho_{c}^{2}/\rho^2}
{\rho^2 \left( 1-\rho_{c}^{2}/\rho^2\right)} \; , \\
\nonumber & & \tilde{g}^{\rho \rho}=1 \; {\rm and}\;  \tilde{g}^{zz}=1 .
\end{eqnarray}
Expressions (\ref{14}) and (\ref{15}) are well defined for $\rho >\rho_c$, i.e.
outside the core region of the vortex flow.

The three vectors $\partial_0$, $\partial_{\varphi}$ and $\partial_z$
define again three Killing vectors $\tilde{\xi}_{(a)}^{\mu}$ of the effective
metric $\tilde{g}_{\mu \nu}$, i.e. 
$\tilde{\nabla}_{\alpha}\tilde{\xi}_{(a)\beta}+\tilde{\nabla}_{\beta}
\tilde{\xi}_{(a)\alpha}=0$. Taking into account the geodesic equations
(\ref{13}), we find that 
$\tilde{k}^{\mu}\tilde{\nabla}_{\mu}(\tilde{\xi}_{(a)}^{\nu}\tilde{k}_{\nu})=0$
and so $\tilde{\xi}_{(a)}^{\mu}\tilde{k}_{\mu}$ are constants of
the light motion and thus we put
\begin{equation}\label{16}
\tilde{k}_0=-E \quad \tilde{k}_{\varphi}=L \quad {\rm and}\quad
\tilde{k}_z=C
\end{equation}
where $E$ $(E>0$, $L$ and $C$ are arbitrary constants, $E$ being 
the energy and $L$ the angular momentum of the light rays. 

\section{Equations of the path of the light }

We parametrize the path of the light $x^{\mu}(\lambda )$ with
the affine parameter $\lambda$ so that we have
\begin{equation}\label{17}
\tilde{k}^{\mu}=\frac{dx^{\mu}}{d\lambda} ,
\end{equation}
obeying the geodesic equations (\ref{13}) for $\rho >\rho_c$.
We limit ourselves to consider trajectories of the light rays
in the plane orthogonal
to the vortex flow. We can put $z(\lambda )=0$ having $C=0$.
We do not write down the geodesic equations (\ref{13}) 
but we know that these are 
differential equations of second order. To determine uniquely the path of 
light, we must know at $\lambda =\lambda_0$ the quantities
$$
\left( x^{0}\right)_0 \; , \;  \left( \frac{dx^0}{d\lambda}\right)_0 \; , \; 
\left( \rho \right)_0 \; , \;
\left( \frac{d\rho}{d\lambda}\right)_0 \; , \;  \left( \varphi \right)_0 \;
{\rm and}\; \left( \frac{d\varphi}{d\lambda}\right)_0 .
$$
However, there exists the constraint (\ref{10}) on these initial data 
and consequently there is five independent initial data.

We consider the characterization of the solutions to the geodesic
equations (\ref{13}) by the constants (\ref{16}). According to (\ref{12}), 
we have the differential equations
\begin{equation}\label{18}
\frac{dx^0}{d\lambda}=-\tilde{g}^{00}E+\tilde{g}^{0\varphi}L \quad {\rm and}
\quad \frac{d\varphi}{d\lambda}=\tilde{g}^{\varphi \varphi}L
-\tilde{g}^{0\varphi}E . 
\end{equation}
The unknown fonction $\rho (\lambda )$ is determined from 
equation (\ref{10}) which can be written
\begin{equation}\label{19}
\left( \frac{d\rho}{d\lambda}\right)^2=
-\tilde{g}^{00}E^2+2\tilde{g}^{0\varphi}EL-\tilde{g}^{\varphi \varphi}L^2 .
\end{equation}
In this case, the path of light is characterized by the five quantities
$$
\left( x^{0}\right)_{0}\; , \; \left( \varphi \right)_0 \; , \;  
\left( \rho \right)_0 \; , \; E \; {\rm and} \; L .
$$
Now we point out that the affine parameter $\lambda$ is only defined up to
a constant factor, therefore we can put $E=1$ by a choice of the
affine parameter.  

We now turn to analyse the differential equation (\ref{19})
for $\rho >\rho_c$. By setting
\begin{equation}\label{20}
x=\frac{\rho}{\rho_c} \; , \; \tau =\frac{\lambda}{\rho_{c}} \; 
{\rm and}\; \gamma =\frac{\rho_c}{L} ,
\end{equation}
equation (\ref{19}) becomes 
\begin{equation}\label{21}
\left( \frac{dx}{d\tau}\right)^2=\frac{x^2}{(x^2-1)\gamma^2}\left[
\left( n^2-\frac{1}{x^2}\right) \gamma^2-\frac{2(-1+n^2)}{x^2}\gamma
-\left( 1-\frac{n^2}{x^2}\right)\frac{1}{x^2}\right] ,
\end{equation}
valid for $x>1$ and $\gamma \not= 0$. We can rewrite (\ref{21}) in the form
\begin{equation}\label{23}
\left( \frac{dx}{d\tau}\right)^2=\frac{n^2x^2-1}{(x^2-1)\gamma^2}\left[
\left( \gamma -\gamma_1(x)\right) \left( \gamma -\gamma_2(x)\right) \right]
\end{equation}
where
\begin{equation}\label{24}
\gamma_1(x)=\frac{n-x}{x(nx-1)} \quad {\rm and}
\quad \gamma_2(x)=\frac{n+x}{x(nx+1)} .
\end{equation}

By virtue of (\ref{18}), the polar angle $\varphi (\tau )$
is then determined by
\begin{equation}\label{30}
\frac{d\varphi}{d\tau}=\frac{1}{x^2-1}\left[ \left( 1-\frac{n^2}{x^2}\right)
\frac{1}{\gamma}-1+n^2\right]
\end{equation}
when $x(\tau )$ is known.

In the case $\gamma =1$, we can easily find the general solution
to equations (\ref{23}) and (\ref{30})
\begin{equation}\label{31}
x(\tau )=\sqrt{1+n^2(\tau_0 -\tau )^2} \quad {\rm and}\quad 
\varphi (\tau )=n\arctan n(\tau_0 -\tau )+\varphi_0 .
\end{equation}

In the case $\gamma \not= 1$, expressions (\ref{23}) and (\ref{30})
have a divergence at $x=1$ and therefore we have
$$
\lim_{x\rightarrow 1}\left( \frac{dx}{d\tau}\right) =\infty \quad {\rm and}
\quad \lim_{x\rightarrow 1}\left( \frac{d\varphi}{d\tau}\right) =\infty .
$$
Since the effective metric (\ref{14}) is singular at $x=1$, we cannot 
discuss the path of the light in terms of $\tau$ at $x=1$.
In any way, the affine parameter $\lambda$ has no physical meaning
since $\tilde{g}_{\mu \nu}$ is an auxiliary metric.
Also, it is convenient to express $x$ in function of $\varphi$ for 
$x>1$ and we obtain
thereby
$$
\lim_{x\rightarrow 1}\left( \frac{dx}{d\varphi}\right) =0 .
$$
This means that the trajectories of the light rays 
arriving at $x=1$ are tangent at the circle $x=1$.   
Then, we can match them with the similar trajectories outgoing at the same
position. 

In consequence, we would directly determine  $x(\varphi )$ by the
following differential equation
\begin{equation}\label{32}
\left( \frac{dx}{d\varphi}\right)^2=\frac{(x^2-1)(n^2x^2-1)}
{(1-n^2)^2[\gamma -\gamma_0(x)]^2}\left[ 
(\gamma -\gamma_1(x))(\gamma -\gamma_2(x))\right]
\end{equation}
where 
\begin{equation}\label{33}
\gamma_0(x)=\frac{x^2-n^2}{x^2(1-n^2)},
\end{equation}
deduced from (\ref{23}) and (\ref{30}).

\section{Qualitative features of the trajectories}

We are now in a position to discuss the qualitative features of the 
trajectories of the light rays. Already for $\gamma =1$, we have the 
particular solution (\ref{31}) which can be rewritten 
\begin{equation}\label{31a}
x(\varphi )=\sqrt{1+\left( \tan \frac{\varphi -\varphi_0}{n}\right)^2} .
\end{equation}
We see that the path of light described by trajectory (\ref{31a})  
arrives at $x=1$ from infinity then goes back at the infinity.

A solution to equation (\ref{32}) exists when its right member is positive.
This fact depends on the value of $\gamma$ with respect to
the roots $\gamma_1(x)$ and $\gamma_2(x)$. The turning points occur when 
$dx/d\varphi =0$. The derivative $dx/d\varphi$ changes of sign there 
except if the derivative of the right member with respect to $\varphi$ 
vanishes also at this point. In this latter case, we have 
a limiting circle. 
Moreover, as proved in the previous section, we see directly from 
equation (\ref{32}) that the trajectories of the light rays  
passing at $x=1$ are tangent at the circle $x=1$.

In order to discuss graphically the form of the trajectories dependent on the
values of $\gamma$, we must draw the curves $\gamma_1$ and $\gamma_2$.
We note that $\gamma_1(1)=1$ and $\gamma_2(1)=1$ and
$\gamma_1(\infty )=0$ and $\gamma_2(\infty )=0$.
The derivatives of $\gamma_1$ and $\gamma_1$  with respect to $x$ are
\begin{equation}\label{25}
\frac{d\gamma_1}{dx}=\frac{n(1-2n+x^2)}{x^2(nx-1)^2} \quad {\rm and}\quad
\frac{d\gamma_2}{dx}=-\frac{n(1+2n+x^2)}{x^2(nx+1)^2} .
\end{equation}
The derivative of the function $\gamma_2$ is always negative for $x>1$
and so $\gamma_2$ is monotonically decresing from 1 to 0. 
On the contrary, the form of the curve of $\gamma_1$ depends on  
whether $n<1$ or $n>1$.

\subsection{Case $n<1$}

The function $\gamma_1$ blows up for $x=1/n$. The derivative 
of the function $\gamma_1$ is always positive. 
For $x>1/n$, we have $\gamma_2(x)>\gamma_1(x)$ and for $1<x<1/n$, we have
$\gamma_2(x)<\gamma_1(x)$. We remark that the component 
$\tilde{g}_{00}$ of the effective metric (\ref{14}) vanishes for 
$\rho =\rho_c/n$ and changes of sign.
\begin{figure}
\begin{picture}(380,380)
\qbezier(100,220)(117,252)(120,366.6)
\qbezier(130,7.6)(133.2,195.8)(350,191.6)
\qbezier(100,220)(157,210.2)(350,206.4)
\put(100,200){\vector(1,0){270}}
\put(100,0){\vector(0,1){370}}
\multiput(125,0)(0,20){19}{\line(0,1){10}}
\put(85,360){$\gamma$}
\put(85,200){0}
\put(85,220){1}
\put(105,210){1}
\put(200,198){$\mid$}
\put(200,210){2}
\put(360,190){$x$}
\put(300,220){$\gamma_2(x)$}
\put(300,175){$\gamma_1(x)$}
\put(130,330){$\gamma_1(x)$}
\put(130,188){$1/n$}
\end{picture}
\caption{Curves $\gamma_1(x)$ and $\gamma_2(x)$ for $n=0.8$}
\end{figure}
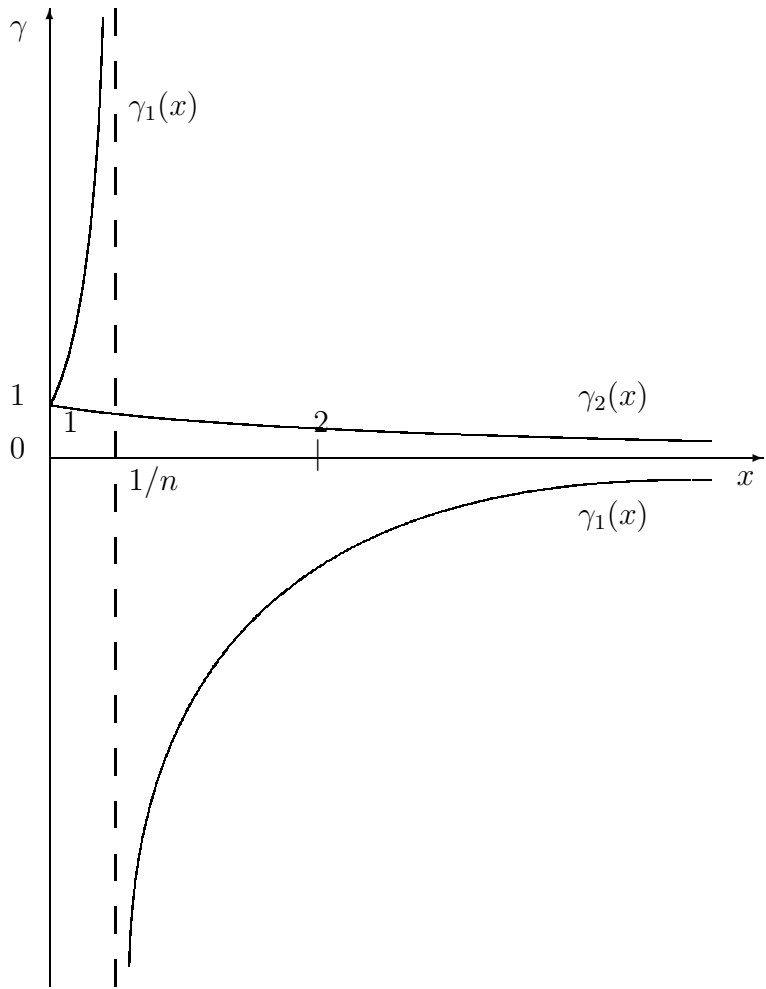

The solution to equation (\ref{32}) exists when
$\gamma >\gamma_1(x)$ or $\gamma <\gamma_1(x)$ for $x>1/n$
and when $\gamma_2(x)<\gamma <\gamma_1(x)$ for $1<x<1/n$ since the
factor in front of $\gamma^2$ in (\ref{32}) is presently negative.
We see graphically on Fig. 1 that for all $\gamma$, $\gamma \not= 1$, 
there is a turning point. We have quasi-hyperbolic trajectories. The
case $\gamma =1$ is the particular solution (\ref{31a}) already studied.

\subsection{Case $n>1$}

The derivative of the function $\gamma_1$ vanishes at 
$x=x_l$ with $x_l=n+\sqrt{n^2-1}$. 
We denote $\gamma_l=\gamma_1(x_l)$ which has the expression
\begin{equation}\label{27}
\gamma_l=-\frac{1}{(n+\sqrt{n^2-1})^2} ,
\end{equation}
satisfying $-1<\gamma_1(x_l)<0$. We have $\gamma_2(x)>\gamma_1(x)$.
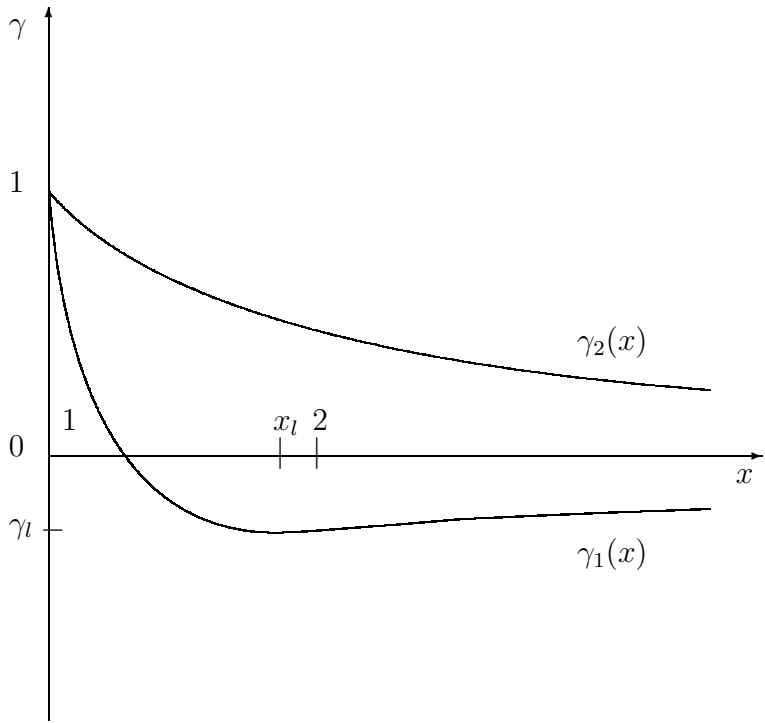
\begin{figure}
\begin{picture}(380,280)
\qbezier(100,200)(110,71)(186,71)
\qbezier(186,71)(195,71)(255,76)
\qbezier(255,76)(270,77)(350,80)
\qbezier(100,200)(151,140)(350,125)
\put(100,100){\vector(1,0){270}}
\put(100,0){\vector(0,1){270}}
\put(85,260){$\gamma$}
\put(85,200){1}
\put(85,100){0}
\put(105,110){1}
\put(200,98){$\mid$}
\put(200,110){2}
\put(360,90){$x$}
\put(300,140){$\gamma_2(x)$}
\put(300,60){$\gamma_1(x)$}
\put(186,98){$\mid$}
\put(185,110){$x_l$}
\put(97,69){$-$}
\put(85,71){$\gamma_l$}
\end{picture}
\caption{Curves $\gamma_1(x)$ and $\gamma_2(x)$ for $n=1.2$}
\end{figure}

The solution to equation (\ref{32}) exists when $\gamma >\gamma_2(x)$ or 
$\gamma <\gamma_1(x)$. We see graphically on Fig. 2 that
when $\gamma >1$ or $\gamma <\gamma_l$, the light reaches tangentially 
to the circle $x=1$ and goes back at the infinity. 
We point out the existence of a limiting circle at $x=x_l$ 
and thus for $\gamma =\gamma_l$ the corresponding light ray spirals 
around this limiting circle. For $\gamma_l<\gamma <1$,
the form of the curves $\gamma_1$ and $\gamma_2$ shows that there are two
turning points. We have either quasi-hyperbolic trajectories or 
quasi-elliptic trajectories, oscillating between $x=1$ and 
$x_{\gamma}$ defined by $\gamma_1(x_{\gamma})=\gamma$.

\section{Conclusion}

In the present work, we have considered that the dielectric vortex
is the relativistic vortex flow of a perfect fluid. The core region 
defined by $\rho <\rho_c$ is not generally considered. Outside the
core region, it is possible to calculate  an
effective metric for which the light rays define a null congruence
of geodesics. By discussing the general features of the trajectories
of the light rays, we have proved that the light rays remain 
external to the core region. The trajectories arriving at $\rho =\rho_c$
are tangent at the circle $\rho =\rho_c$. However, we have merely an
idealized model of dielectric vortex and, since the effective metric
is singular at $\rho =\rho_c$, we cannot determine the
behaviour of the light within geometrical optics in the neighbourhood
of $\rho =\rho_c$.


\begin{thebibliography}{99}

\bibitem{le1} Leonhardt, U and Piwnicki, P (1999) {\em Phys. Rev.} A 
{\bf 60}, 4301.
\bibitem{vi} Visser, M (2000) gr-qc/0002011.
\bibitem{le2} Leonhardt, U and Piwnicki, P (2000) {\em Phys. Rev. Lett.} 
{\bf 84}, 822. 
\bibitem{bo} Boisseau, B (2000) {\em Phys. Rev.} D {\bf 61}, 083504.
\bibitem{ca} Carter, B and Langlois, D (1995) {\em Phys. Rev.} D {\bf 52},
4640. 
\bibitem {go} Gordon, W (1923) {\em Ann. Phys. (Leipzig)} {\bf 72}, 421.
\bibitem{ph} Pham Mau Quan (1956) {\em C. R. Acad. Sci. (Paris)} {\bf 242},
465.
\bibitem{to} Tourrenc, Ph (1981) {\em J. Phys. Colloq.} {\bf 8}, 441.

\end{thebibliography}
\end{document}